\documentclass[aps,pra,twocolumn,showpacs,superscriptaddress]{revtex4-2}

\usepackage{dcolumn}
\usepackage{bm}
\usepackage{amssymb}
\usepackage{subfigure}
\usepackage{here}
\usepackage{ulem}
\usepackage{enumerate}
\usepackage{cancel}
\usepackage{flushend}
\usepackage{bbold}
\usepackage{mathtools}
\usepackage{float}
\usepackage{stmaryrd}
\usepackage{colortbl}
\usepackage[table]{xcolor}
\usepackage{array,ragged2e}
\usepackage{amsmath,amssymb,amsfonts,stmaryrd,wasysym,graphicx,multirow,color,textcomp,subfigure}
\usepackage{url}
\usepackage[colorlinks=true,citecolor=blue,urlcolor=blue]{hyperref}
\usepackage{romannum}
\usepackage{tabularx}
\usepackage{changes}

\usepackage[utf8]{inputenc}
\usepackage{tikz-cd}

\usepackage{graphicx}
\usepackage{color}

\newcommand{\realb}[1]{\text{Re}[$#1$]}

\newcommand{\eq}[1]{Eq.~(\ref{#1})}

\newenvironment{helv}{\fontfamily{phv}\selectfont}{\par}
\DeclareTextFontCommand{\texthelv}{\helv}

\makeatletter
\newcommand{\thickhline}{%
    \noalign {\ifnum 0=`}\fi \hrule height 1pt
    \futurelet \reserved@a \@xhline
}
\newcolumntype{"}{@{\hskip\tabcolsep\vrule width 1pt\hskip\tabcolsep}}
\makeatother

\newcommand{\figsizeone}{0.9}

\begin{document}

\draft
\title{Realization of geometric phase topology induced by multiple exceptional points}

\author{Jung-Wan Ryu}
\address{Center for Theoretical Physics of Complex Systems, Institute for Basic Science (IBS), Daejeon 34126, Republic of Korea.}
\author{Jae-Ho Han}
\email{jaehohan@kaist.ac.kr}
\address{Department of Physics, Korea Advanced Institute of Science and Technology (KAIST), Daejeon 34141, Republic of Korea.}
\author{Chang-Hwan Yi}
\email{yichanghwan@hanmail.net}
\address{Center for Theoretical Physics of Complex Systems, Institute for Basic Science (IBS), Daejeon 34126, Republic of Korea.}
\date{\today}

\begin{abstract}
Non-Hermitian systems have Riemann surface structures of complex eigenvalues that admit singularities known as exceptional points. Combining with geometric phases of eigenstates gives rise to unique properties of non-Hermitian systems, and their classifications have been studied recently. However, the physical realizations of classes of the classifications have been relatively limited because a small number of modes and exceptional points are involved. In this work, we show in microcavities that all five classes [J.-W. Ryu, {\it et al.,} Commun. Phys. {\bf 7}, 109 (2024)] of three modes can emerge with three exceptional points. In demonstrations, we identified various combinations of exceptional points within a two-dimensional parameter space of a single microcavity and defined five distinct encircling loops based on three selected exceptional points. According to the classification, these loops facilitate different mode exchanges and the acquisition of additional geometric phases during the adiabatic encircling of exceptional points. Our results provide a broad description of the geometric phases-associated topology induced by multiple exceptional points in realistic physical systems.
\end{abstract}

\maketitle

\section{Introduction}

Non-Hermitian systems can have complex eigenvalues and non-orthogonal eigenstates \cite{moiseyev2011non, El-Ganainy2018non, Ashida2020non, Bergholtz2021exceptional}. The complex eigenvalues generate Riemann surfaces for the multivalued complex function, leading to rich and complicated phenomena that differ from Hermitian systems with real eigenvalues and orthogonal eigenstates. One of the most prominent features of non-Hermitian systems is the existence of non-trivial singularities, known as exceptional points (EPs), where both eigenvalues and eigenstates coalesce \cite{Kato1976perturbation, Heiss1990avoided, Lee2008divergent, Ryu2009coupled, Dietz2011exceptional, Ding2015coalescence, Ding2016emergence, Cui2019exceptional}. For a homotopy classification of multiple eigenstates, one considers an encircling loop in parameter space and the eigenstate exchange effect following it, which depends on the Riemann surface structures inside the loop \cite{Cartarius2009exceptional, Ryu2012analysis, Zhong2018winding, Pap2018non, Luitz2019exceptional, Hu2021knots, Patil2022measuring, Koenig2023braid}. State exchange occurs in the presence of EPs inside the loops, and the eigenstate exchange effect can be described by the equivalence classes of the permutation group that characterizes the multiple EPs \cite{Wojcik2020homotopy, Li2021homotopical, Ryu2022classification, Zhong2023numerical}.

Adiabatic encircling of EPs leads to an additional geometric phase along with an exchange of eigenstates \cite{Heiss1999phases, Heiss2000repulsion, Dembowski2001experimental, Dembowski2004encircling, Lee2010geometric, Ryu2012geometric}. Studying these phenomena, an elaborated classification framework for EPs has been proposed. Recently, it was clarified in \cite{Ryu2024exceptional} that incorporating the eigenstate exchange effect and the additional phases results in finer classifications depending on the presence of an additional geometric phase of $\pi$ after a double encircling of EPs. The additional geometric phases in non-Hermitian band systems are imprinted in the eigenstates after encircling EPs along a closed loop in momentum space, manifesting as the Berry phase in two-dimensional systems or the Zak phase in one-dimension \cite{Thouless1982quantized, Sheng2003phase, Hasan2010colloquium, Qi2011topological, Zak1989berry, Atala2013direct, Xiao2014surface1, Xiao2014surface2}. On the other hand, the additional phases arise in two-dimensional cavity systems from continuous changes in the wavefunctions, measured along a closed loop that encircles EPs in parameter space. To date, numerous studies have been conducted on this topic based on optical resonator platforms \cite{Dembowski2001experimental, Dembowski2004encircling, Lee2010geometric, Ryu2012geometric}. However, in most of those physical realizations, only one or two EPs have been focused on, and the classifications have not been deemed significantly helpful.

In this work, we demonstrate that multiple EPs and all topological classes based on complex energy structures with additional geometric phases \cite{Ryu2024exceptional} can emerge in microcavities. To identify all these classes, it is necessary to find pairs of EPs connected by branch cuts. However, locating such pairs in strongly deformed chaotic microcavities is challenging, as the deformations can separate them significantly, making it difficult to find pairs within a tractable parameter range. We address this issue by using optical modes in deformed elliptic cavities, where paired EPs are located near each other. We focus on three modes associated with three EPs: a pair of EPs and a single EP. In the following, we discuss the five topologically distinct classes, each characterized by different encircling loops in parameter space.

\section{Results}

\subsection{Five classes for three eigenmodes}

\begin{figure*}[t]
\begin{center}
\includegraphics[width=1.5\columnwidth]{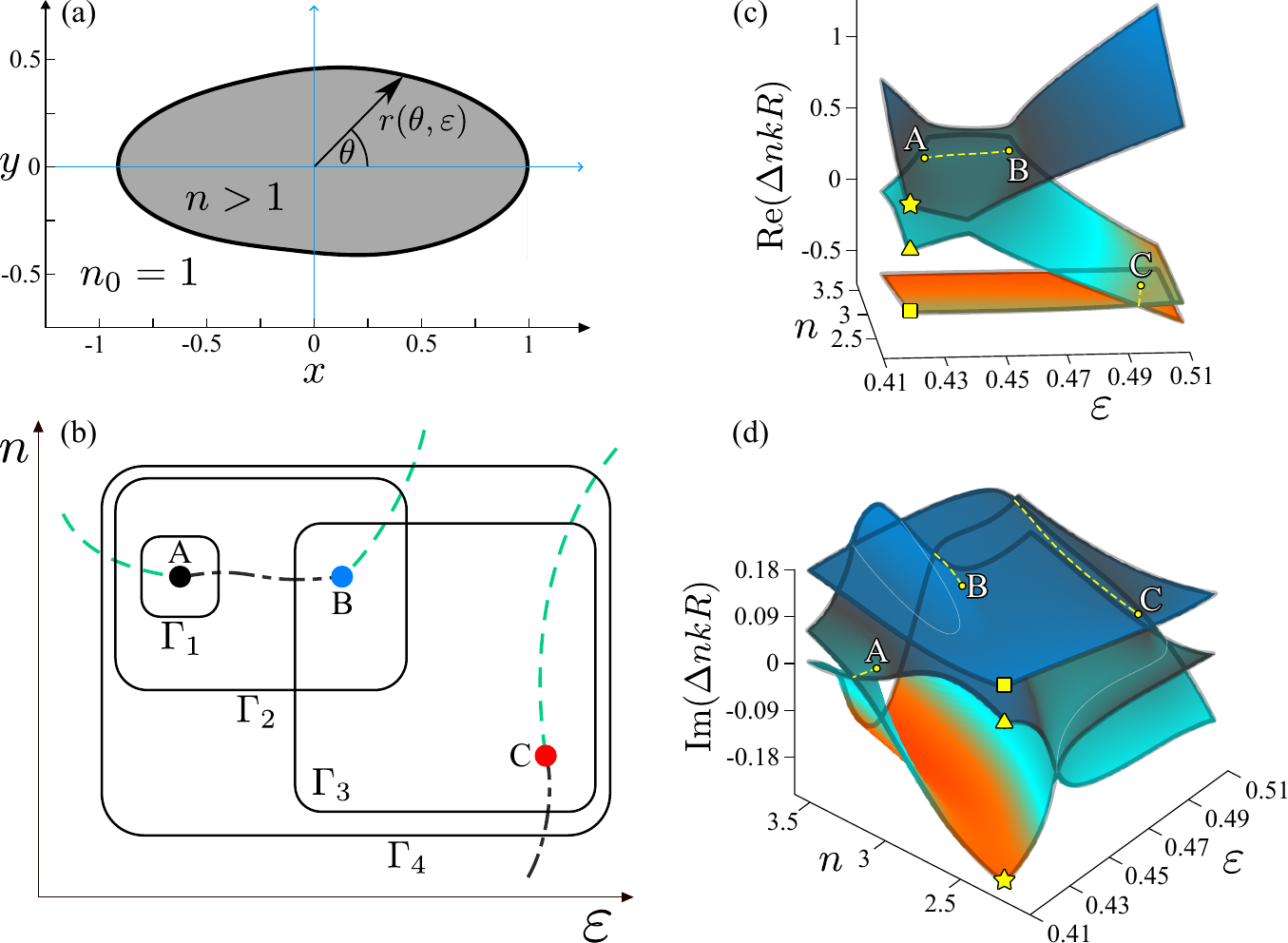}
\caption{(a) Dielectric microdisk system configuration with a deformed elliptic boundary shape as given in~\eq{eq:system}. (b) Three EP positions (A, B, and C) and four encircling loops, $\Gamma_1$, $\Gamma_2$, $\Gamma_3$, and $\Gamma_4$ examined. Black dash-dotted and green dashed lines represent real and imaginary branch cuts. Riemann surfaces of (c) real and (d) imaginary parts of complex eigenfrequencies of three eigenmodes. Yellow dashed lines denote associated real and imaginary branch cuts. Note that the eigenfrequencies in (c) and (d) are presented relative to the average value of the three modes for visual clarity: $\Delta nk_i(\varepsilon,n)R=nk_i(\varepsilon,n)R-\langle nk(\varepsilon,n)R \rangle$ where $\langle nk(\varepsilon,n)R \rangle=1/3\sum_{j=1}^3nk_j(\varepsilon,n)R$}
\label{fig1}
\end{center}
\end{figure*}

We examine three eigenmodes in deformed microcavities. The corresponding eigenfrequencies of these modes are complex-valued, with the imaginary parts corresponding to the decay rates of the modes. The decay rate of modes originates from the outgoing boundary conditions at infinity and characterizes the system as non-Hermitian \cite{Tuereci2005modes, Cao2015dielectric}. Details will be explained in Sec.~\ref{Cavity}. The complex eigenfrequencies in the two-dimensional parameter space are generally non-separable because of EPs. Here, non-separability implies that modes switch continuously in response to the adiabatic parameter variation. The Riemann surfaces of complex eigenfrequencies in Figs.~\ref{fig1}(c) and (d) demonstrate the three non-separable modes associated with three EPs.

Considering the complex eigenfrequencies of three modes, there can be three different topological structures of the Riemann surfaces of eigenfrequencies depending on the encircling loops around EPs: three one-cycle modes ($1^3$), one one-cycle and one two-cycle mode ($1^1 2^1$), and one three-cycle mode ($3^1$) \cite{Ryu2022classification}.
By combining these three classes with additional topological phases of the eigenmodes, five different classes emerge \cite{Ryu2024exceptional}:
\begin{enumerate}
    \item $(1^3)$; three one-cycle modes without additional phases
    \item $(1^1 \bar{1}^2)$; one one-cycle mode without an additional phase and two one-cycle modes with additional phases of $\pi$
    \item $(1^1 \bar{2}^1)$; one one-cycle mode without an additional phase and one two-cycle mode with an additional phase of $\pi$
    \item $(\bar{1}^1 2^1)$; one one-cycle mode with an additional phase of $\pi$ and one two-cycle mode without an additional phase
    \item $(3^1)$; one three-cycle mode without an additional phase
\end{enumerate}
The classes and associated examples of mode exchanges are summarized in Table~\ref{table3}. These five classes correspond to the topological structures of multi-bands and the associated Berry phases in the three-band model \cite{Ryu2024exceptional}. 

In the following, we will demonstrate these five classes for three eigenmodes in a single optical microcavity, providing a practical application of our mathematical idea in a realistic physical system.

\subsection{Deformed elliptic cavities and EPs}
\label{Cavity}

An optical microcavity consists of a dielectric material which is defined inside a slightly deformed elliptical shape [Fig.~\ref{fig1}(a)]. The angle-dependent radius of the boundary shape is given in polar coordinates as,
\begin{equation}\label{eq:system}
r(\theta,\varepsilon) = \frac{\varepsilon R}{ \sqrt{1-(1-{\varepsilon}^2) \cos^2(\theta)}}+ {\delta R} \cos(\theta-\theta_0)\ ,
\end{equation}
where $\varepsilon$ is the aspect ratio of an unperturbed ellipse, and $\delta$ and $\theta_0$ are additional deformation parameters breaking the geometric symmetry of the ellipse. The parameter set $(\varepsilon,\delta)=(1,0)$ corresponds to a circle with a radius $R$.

Given the boundary shape of \eqref{eq:system}, we obtain the optical eigenmodes by solving the two-dimensional Maxwell equations reduced to the Helmholtz wave equation,
\begin{align}
-\nabla^2\vec{\psi}(\mathbf{r})=n^2(\mathbf{r})k^2\vec{\psi}(\mathbf{r})\ ,
\label{eq:helm}
\end{align}
imposing a transverse-magnetic [TM; $\vec{\psi}=(0,0,E_z)$] dielectric boundary condition at the cavity--vacuum interface \cite{Jackson1999classical, Chang1996optical}. For $|\mathbf{r}| \to \infty$, we apply the pure-outgoing wave boundary condition. This condition results in the wavenumber $k = \omega/c$ and the time-harmonic frequency $\omega$ being complex-valued, i.e., $k, \omega \in \mathbb{C}$. Here, $ c $ is the speed of light, and $ i^2 = -1 $. The refractive index $ n(\mathbf{r}) $ is piecewise constant, with $ n > 1 $ inside the microdisk and $ n_0 = 1 $ outside. Note that, for simplicity, we will henceforth use $ \psi $ to represent the scalar electric field $ E_z $, instead of the vector notation $ \vec{\psi} $. Additionally, the wavenumber will be expressed in the dimensionless form $ kR $ for clarity, and since $nkR=n\omega R/c$, we will call $nkR$ eigenfrequency throughout the paper for convenience.
\begin{table}
\caption{A notation of classes, examples of mode exchanges for three eigenmodes, and combinations of EPs. $C^j$ represents ``$j$ $C$-cycle" modes, and the bar represents an additional phase $\pi$ of each mode. The right arrow denotes an encircling EPs, and the minus signs represent the additional phase $\pi$ after the encircling EPs.}
\renewcommand{\arraystretch}{1.5}
\begin{center}
\scalebox{1.0}{%
\begin{tabular}{ | c | p{30mm} | p{30mm} | }
\hline
Class & Mode exchanges & EPs\\
\hline
$1^{3}$ & $1 \rightarrow 1$ \newline $2 \rightarrow 2$ \newline $3 \rightarrow 3$ & No EP \\
\hline
$1^1 \bar 1^{2}$ & $1 \rightarrow -1$ \newline $2 \rightarrow -2$ \newline $3 \rightarrow 3$ & A pair of EPs \newline [Sec.~\ref{loop2}] \\
\hline
$1^{1} \bar 2^{1}$ & $1 \rightarrow -2 \rightarrow -1$ \newline $2 \rightarrow 1 \rightarrow -2$ \newline $3 \rightarrow 3$ & One EP \newline [Sec.~\ref{loop1}] \\
\hline
$\bar 1^{1} 2^{1}$ & $1 \rightarrow -1$ \newline $2 \rightarrow -3 \rightarrow 2$ \newline $3 \rightarrow -2 \rightarrow 3$ & Three EPs \newline [Sec.~\ref{loop4}]\\
\hline
$3^{1}$ & $1 \rightarrow 3 \rightarrow 2 \rightarrow 1$ \newline $2 \rightarrow 1 \rightarrow 3 \rightarrow 2$ \newline $3 \rightarrow 2 \rightarrow 1 \rightarrow 3$ & Two intersected EPs \newline [Sec.~\ref{loop3}]\\
\hline
\end{tabular}}
\end{center}
\label{table3}
\end{table}

\subsection{Encircling loops and topological classes}

We obtain three eigenmodes in the deformed elliptic cavity, which exhibit Riemann surfaces for the real and imaginary parts of the complex eigenfrequencies [Figs.~\ref{fig1}(c) and (d)]. In the parameter space $ (\varepsilon, n) $, with fixed asymmetric parameters $ (\delta, \theta_0) = (0.047, \pi/5) $, three EPs are associated with these modes [Fig.~\ref{fig1}(b)]. The modes are labeled as the first, second, and third based on the order of the real parts of their complex eigenfrequencies. The first and second modes are connected by EP$_\mathrm{A}$ and EP$_\mathrm{B}$, while the second and third modes are connected by EP$_\mathrm{C}$.

The EP$_\mathrm{A}$ and EP$_\mathrm{B}$ are a pair of EPs sharing two associated modes (the first and second modes). Near each EP, the complex energy structure can be described by the form,
\begin{align}
E = \sqrt{z-z_{\text{EP}_i}}\ \text{or}\ \sqrt{(z-z_{\text{EP}_i})^*}\ ,\ i=\text{A, B, C},
\end{align}
depending on the vorticity of the structure~\cite{Shen2018vortex}. Here, $z = \epsilon + ni$ is the complex number corresponding to the two-dimensional parameter space point $(\epsilon, n)$, and $z_{\text{EP}_i}$ are the points of EPs. An asterisk (*) represents the complex conjugate. The two EPs, EP$_\mathrm{A}$ and EP$_\mathrm{B}$, have different vorticities, allowing a real branch cut to connect the EPs. EP$_\mathrm{C}$ shares the second mode with EP$_\mathrm{A}$ and EP$_\mathrm{B}$ and has non-trivial effects on the mode exchange, although the real branch cut is not connected to EP$_\text{A}$ or EP$_{\rm B}$. In Fig.~\ref{fig1}(b), the configuration of EPs and branch cuts is shown. The real and imaginary eigenfrequency structures in the parameter space are shown in Figs.~\ref{fig1} (c) and (d), respectively.

Now, we explicitly examine the four encircling loops, $\Gamma_i$, $i=1,2,3,4$ [Fig.~\ref{fig1} (b)]: The loop $\Gamma_1$ encircles only EP$_\mathrm{A}$, and the loop $\Gamma_2$ encircles two EPs, EP$_\mathrm{A}$ and EP$_\mathrm{B}$. The loop $\Gamma_3$ encircles EP$_\mathrm{B}$ and EP$_\mathrm{C}$, and the loop $\Gamma_4$ encircles all three EPs. We omit the trivial case of the loop, which does not encircle any EPs, corresponding to class $1^3$. In the following subsections, we calculate the change of states and Berry phases after the adiabatic encircling along loops ($\Gamma_i$s) in the parameter space and discuss the classes of the mode exchanges.

\subsubsection{Loop $\Gamma_1$: Single EP}
\label{loop1}

\begin{figure}[t]
\begin{center}
\includegraphics[width=\figsizeone\columnwidth]{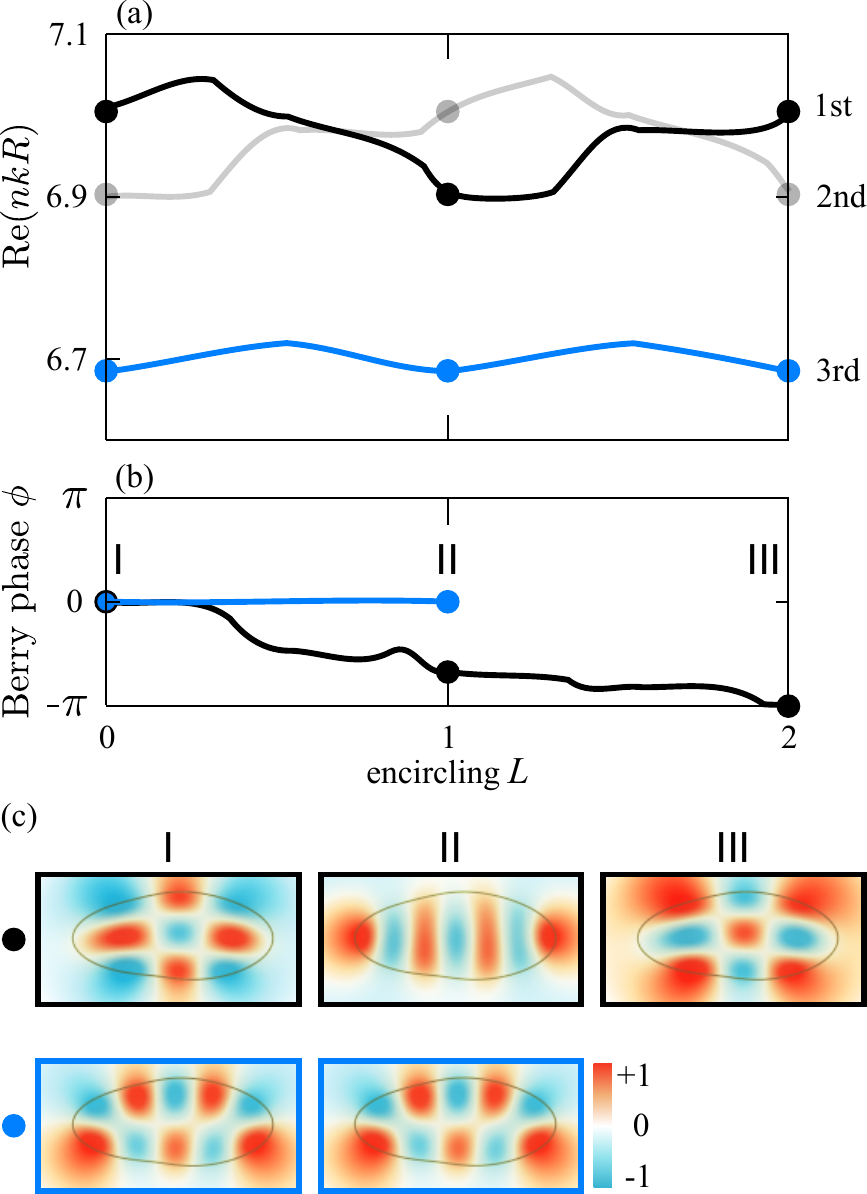}
\caption{(a) The real parts of complex eigenfrequencies and (b) additional geometric phases when encircling an EP along the loop $\Gamma_1$. (c) shows \realb{\psi(\mathbf{r})}, the spatial distributions of the real part of eigenmodes normalized by max$\{$\realb{\psi(\mathbf{r})}$\}$. The red and blue colors represent positive and negative values, respectively. The first mode changes into the second mode after the first encircling. The second mode changes into the first mode after the second encircling. Finally, the mode obtains additional geometric phase $\pi$. The third mode does not change in spite of encircling the EP.}
\label{fig2}
\end{center}
\end{figure}

In this section, the loop $\Gamma_1$ encircling a single EP (EP$_\mathrm{A}$) is analyzed [see Fig.~\ref{fig1}(b)]. The first and second modes switch after the first encircling of the EP and return to themselves with an additional geometric phase of $\pi$ after the second encircling [Fig.~\ref{fig2}]. The additional geometric phases are obtained from the total variation of the biorthogonal inner products of eigenmodes during an adiabatic encircling of the EPs \cite{Garrison1988complex, Mailybaev2005geometric, Cui2012geometric, Wagner2017numerical}. In Fig.~\ref{fig2}, the opposite colors (signs: $\pm$) of the two wave patterns of the first modes when $L=0$ and $L=2$ represent the additional phase of $\pi$, where $L$ is the number of encirclings. The third mode does not change, and there is no additional geometric phase after the first encircling. We can confirm this by observing the same colors of the two wave patterns of the third mode when $L=0$ and $L=1$. As a result, the topological class of the three modes is $1^1 \bar{2}^1$, denoting one separable mode without an additional phase and two non-separable modes with an additional phase of $\pi$. Note that encircling different EPs, such as EP$_\mathrm{B}$ or EP$_\mathrm{C}$, corresponds to the identical class as well.

\subsubsection{Loop $\Gamma_2$: Two EPs - a pair of EPs}
\label{loop2}

\begin{figure}[t]
\begin{center}
\includegraphics[width=\figsizeone\columnwidth]{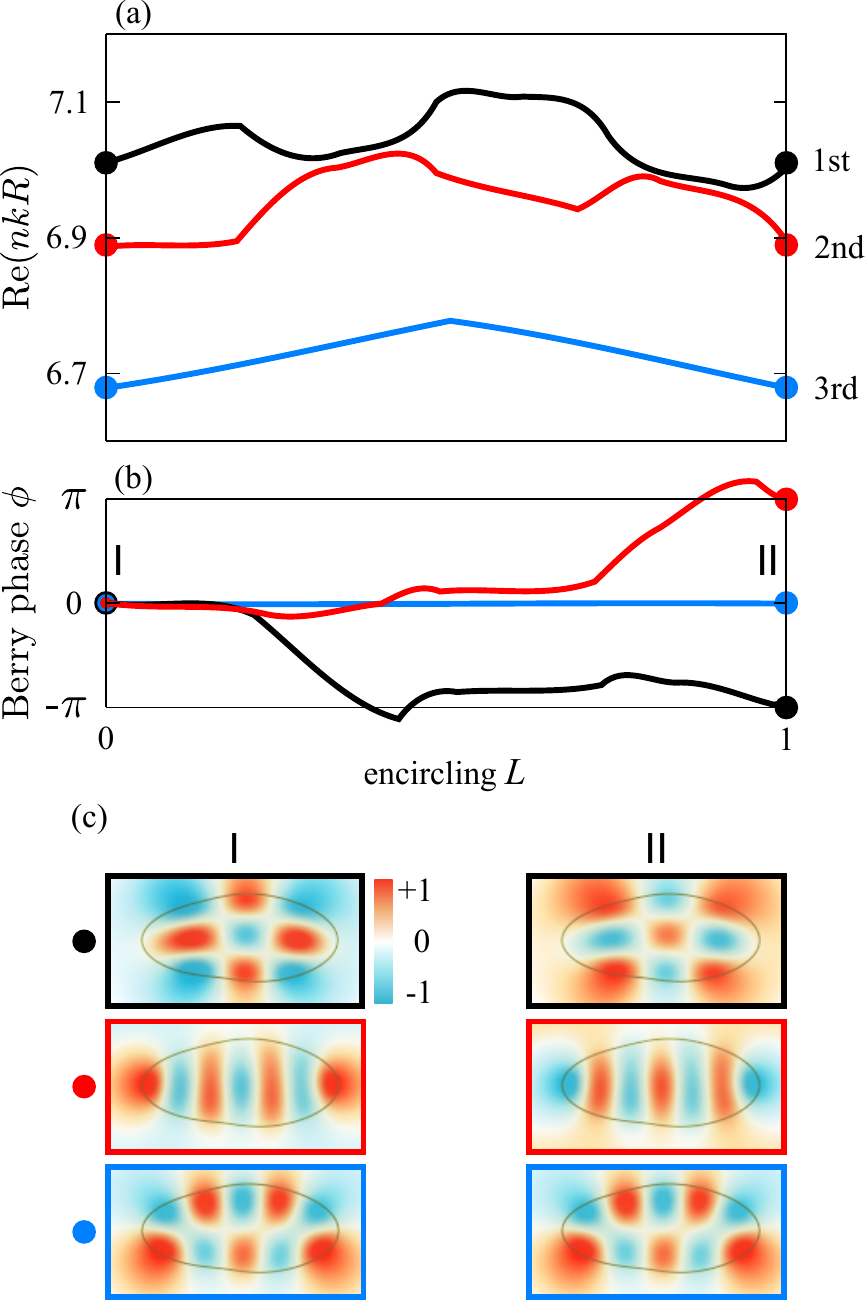}
\caption{(a) The real parts of complex eigenfrequencies and (b) additional geometric phases when encircling a pair of EPs along the loop $\Gamma_2$. (c) shows \realb{\psi(\mathbf{r})} normalized by max$\{$\realb{\psi(\mathbf{r})}$\}$. Three modes do not change, but two modes obtain additional phases $\pi$, and one mode does not obtain additional phases.}
\label{fig3}
\end{center}
\end{figure}

The loop $\Gamma_2$ encircles the two paired EPs, EP$_\mathrm{A}$ and EP$_\mathrm{B}$ [see Fig.~\ref{fig1}(b)], which share two associated eigenmodes, the first and second modes. In this case, we can observe that all three modes do not change after encircling a pair of EPs. However, the first and second modes obtain additional phases of $\pi$, while the third mode does not obtain the additional phase after the encircling [Fig.~\ref{fig3}]. The opposite colors of the two wave patterns of the first and second modes when $L=0$ and $L=1$ represent the additional phase of $\pi$, and the same colors of the two wave patterns of the third mode when $L=0$ and $L=1$ represent no additional phase. It implies that the topological class of the three modes here is $1^1 \bar{1}^2$, denoting one separable mode without an additional phase and two separable modes with an additional phase of $\pi$.

\subsubsection{Loop $\Gamma_3$: Two EPs - intersected EPs}
\label{loop3}

\begin{figure}[t]
\begin{center}
\includegraphics[width=\figsizeone\columnwidth]{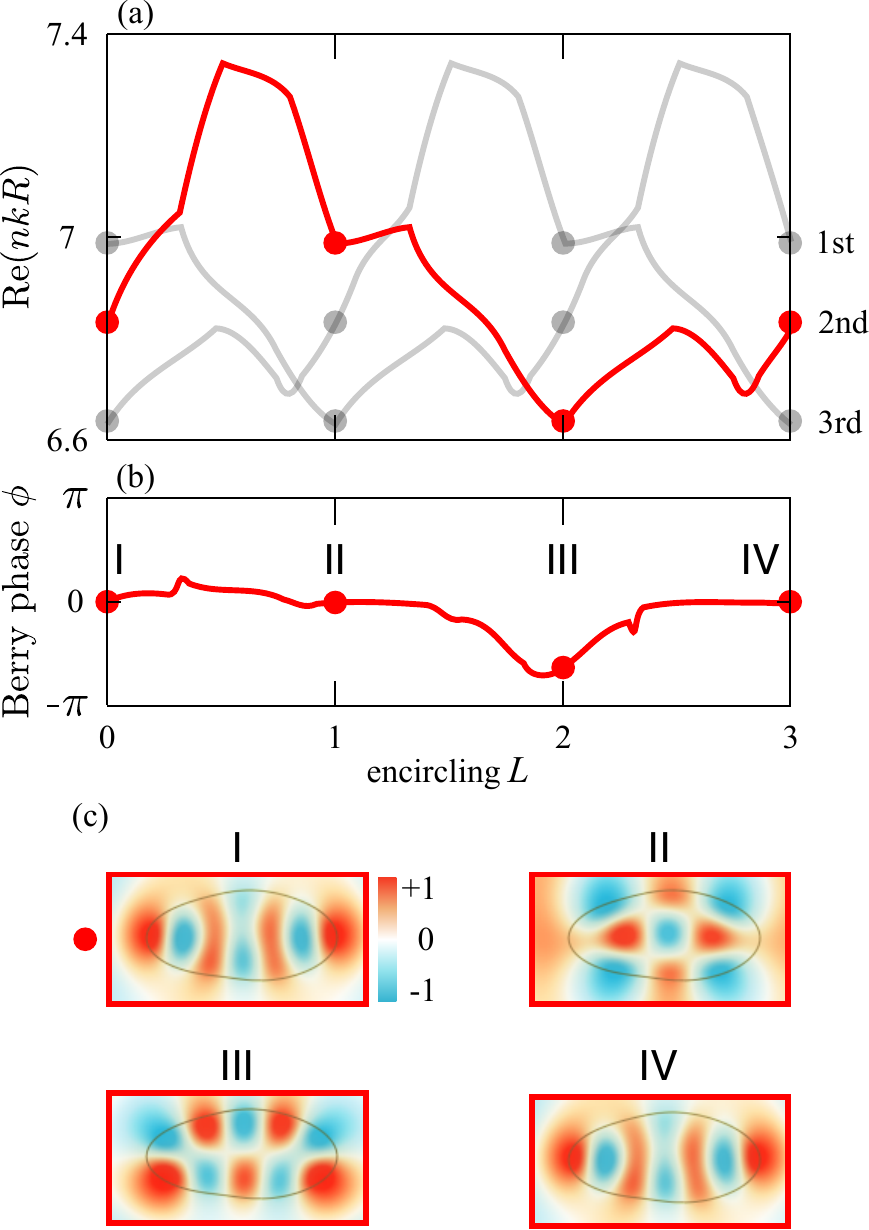}
\caption{(a) The real parts of complex eigenfrequencies and (b) additional geometric phases when encircling two intersected EPs along the loop $\Gamma_3$. (c) shows \realb{\psi(\mathbf{r})} normalized by max$\{$\realb{\psi(\mathbf{r})}$\}$. The second mode changes into the first mode after the first encircling, and the first mode changes into the third mode after the second encircling. Finally, the third mode changes into the second mode after the third encircling. The mode does not obtain additional geometric phase.}
\label{fig4}
\end{center}
\end{figure}

The loop $\Gamma_3$ encircles two intersected EPs, EP$_\text{B}$ and EP$_\mathrm{C}$ [see Fig.~\ref{fig1}(b)], which share one associated eigenmode, the second mode. Here, we can observe that all three modes are not separable and thus the modes always change after encircling the two EPs [Fig.~\ref{fig4}]. For example, if we start from the second mode, it changes into the first mode after the first encircling of the EPs, then into the third mode after the second encircling, and finally back to the second mode after the third encircling. The same colors of the two wave patterns of the second mode when $L=0$ and $L=3$ represent no additional phase. Therefore, the class of the three modes is $3^1$, denoting one non-separable three-cycle mode without an additional phase. We can readily verify that the results are equivalent regardless of the starting modes.

\subsubsection{Loop $\Gamma_4$: Three EPs}
\label{loop4}

\begin{figure}[t]
\begin{center}
\includegraphics[width=\figsizeone\columnwidth]{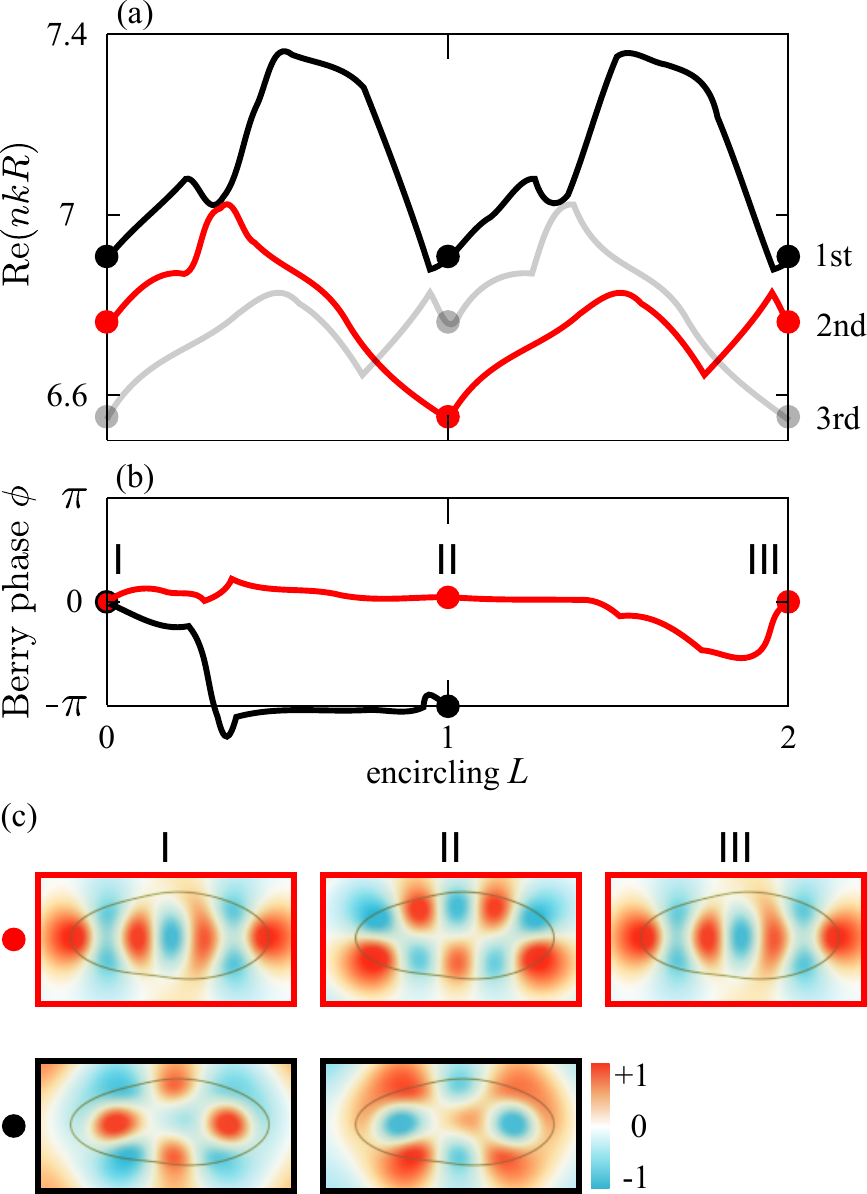}
\caption{(a) The real parts of complex eigenfrequencies and (b) additional geometric phases when encircling three EPs along the loop $\Gamma_4$. (c) shows \realb{\psi(\mathbf{r})} normalized by max$\{$\realb{\psi(\mathbf{r})}$\}$. The second mode changes into the third mode after the first encircling, and the third mode changes into the second mode after the second encircling. The mode does not obtain the additional geometric phase. The first mode does not change but obtains the additional phase $\pi$.}
\label{fig5}
\end{center}
\end{figure}

Lastly, we consider the loop $\Gamma_4$ encircling three EPs, EP$_\mathrm{A}$, EP$_\mathrm{B}$, and EP$_\mathrm{C}$ [see Fig.~\ref{fig1}(b)]. The second and third modes are switched after the first encircling of the EPs and return to themselves without an additional geometric phase after the second encircling [Fig.~\ref{fig5}]. The same colors of the two wave patterns of the second mode when $L=0$ and $L=2$ represent no additional phase. The first mode does not change, but there is an additional geometric phase of $\pi$ after the first encircling, which is represented by the opposite colors of the two wave patterns of the first mode when $L=0$ and $L=1$. As a result, the topological class of the three modes turns out to be $\bar{1}^1 2^1$, denoting one separable mode with an additional phase of $\pi$ and two non-separable modes without an additional phase. It is emphasized that the two classes, $1^1 \bar{1}^2$ on loop $\Gamma_2$ and $\bar{1}^1 2^1$ on loop $\Gamma_4$, can be obtained only when the encircling loop embeds the paired EPs (EP$_\mathrm{A}$ and EP$_\mathrm{B}$ in our demonstration).

\section{Summary}

We have demonstrated all five topological classes by combining Riemann surface structures with the additional geometric phases of three eigenmodes in deformed elliptic microcavities. In a single microcavity, we presented two different types of combinations of two exceptional points. A pair of exceptional points shares two associated eigenmodes, while two intersected exceptional points share only one associated mode. Encircling different combinations of exceptional points enables modes to exchange in distinct ways and acquire additional geometric phases. These mode exchanges and additional geometric phases are well described by five topological classes for three eigenmodes, depending on the encircling of multiple EPs. As the number of modes and associated exceptional points increases, the topology related to mode exchanges and additional geometric phases becomes more complex but can still be understood through the topological concepts of homotopy and permutation groups.

\section*{acknowledgments}
We acknowledge financial support from the Institute for Basic Science in the 
Republic of Korea through the project IBS-R024-D1.

%

\end{document}